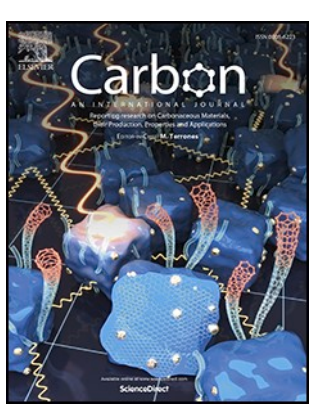

# Nanoscale graphitization and defect evolution in silicon-vacancy center-containing nanodiamonds under high-pressure high-temperature annealing

M. De Feudis [a,b,*], B. Yavkin [c], L. Henry [d], K.O. Ho [c], M.-P. Adam [c], P. Goldner [a], F. Bénédic [e], S. Desgreniers [f], J.-F. Roch [c]

[a] Institut de Recherche de Chimie Paris, Chimie ParisTech, UMR CNRS 8247, PSL Research University, Paris, 75005, France
[b] CY Cergy Paris Université, Cergy-Pontoise, 95031, France
[c] Université Paris-Saclay, CNRS, ENS Paris-Saclay, CentraleSupelec, LuMIn, Gif-sur-Yvette, F-91190, France
[d] Synchrotron SOLEIL, L'Orme des Merisiers, Saint-Aubin, Gif-sur-Yvette, 91192, France
[e] Laboratoire des Sciences des Procédés et des Matériaux, UPR CNRS 3407, Université Sorbonne Paris Nord, Villetaneuse, 93430, France
[f] Laboratoire de Physique des Solides Denses, Department of Physics, University of Ottawa, Ottawa, ON, K1N 6N5, Canada

ABSTRACT

Group-IV color centers, such as the silicon-vacancy (SiV) defect, are highly promising for solid-state quantum technologies. However, nanodiamonds typically exhibit significant lattice strain and structural disorder, which degrade their optical properties and hinder the resolution of the fine spectral structure at cryogenic temperatures. High-pressure high-temperature (HPHT) annealing offers a potential route to relax internal strain, although the phase stability of diamond at the nanoscale under such conditions remains poorly constrained. Here, we investigate the structural evolution of nanodiamonds during HPHT annealing using a Paris–Edinburgh press coupled with *in situ* synchrotron X-ray diffraction at SOLEIL. A dedicated sample assembly combining nanodiamonds – NaCl – Pt enabled accurate pressure–temperature calibration and real-time monitoring of phase transformations. The diffraction data reveal that the onset of diamond-to-graphite transition occurs at approximately 1800 K at 2 GPa and 2120 K at 4 GPa under the applied HPHT heating protocol. These experimentally determined graphitization onsets define a practical pressure–temperature processing window for HPHT annealing of nanodiamonds while avoiding detectable graphitization and provide a calibrated framework for reliable off-beam annealing treatments that avoid graphitization. Photoluminescence measurements on samples annealed below the graphitization threshold show improved optical response, with partial resolution of the SiV fine structure at 12 K. These optical measurements suggest a relationship between nanoscale phase stability and the optical response of individual SiV-containing nanodiamonds following HPHT annealing. The experimentally established HPHT processing window provides a practical framework for the controlled processing of quantum nanodiamonds while avoiding graphitization.

## 1. Introduction

Color centers in diamond are key solid-state platforms for quantum technologies, enabling optical initialization, control, and readout of spin states for applications in communication, information processing, and sensing. Group-IV vacancy (G4V) centers, and in particular the silicon-vacancy (SiV) defect, offer a strong zero-phonon line emission (~80% at room temperature), inversion symmetry, and narrow, stable optical transitions [1]. At cryogenic temperature, the SiV center exhibits a characteristic four-line fine structure arising from spin–orbit and Jahn–Teller interactions, which is highly attractive for indistinguishable photon generation, quantum memories, and coherent spin control [2–6]. Integrating SiV centers into nanodiamonds (NDs) opens opportunities for nanoscale photonics, bioimaging, thermometry, and sensing under extreme environments such as high pressure and magnetic field [7–12]. Recent advances have demonstrated the possibility of incorporating SiV centers into NDs over a broad range of particle sizes, including ultrasmall detonation NDs containing only a few SiV emitters per particle and larger chemical vapour deposition (CVD)-derived NDs optimized for quantum photonic applications [13,14]. Several synthesis routes are currently available for the production of nanodiamonds, including detonation synthesis, milling of bulk high-purity high-pressure high-temperature (HPHT) diamond, direct HPHT synthesis from molecular precursors, and CVD. Depending on the synthesis route, nanodiamonds may exhibit significant differences in particle size distribution, morphology, surface chemistry, defect incorporation, and optical properties. For example, HPHT nanodiamonds synthesized from molecular precursors have been shown to host SiV centers with exceptionally narrow optical linewidths, while CVD-grown nanodiamonds provide a versatile platform for the controlled incorporation of color centers and engineering of crystal growth conditions [15–22]. Direct HPHT synthesis from molecular precursors has also emerged as a route for producing color-center-containing nano- and microdiamonds, including SiV-, GeV-, and SnV-doped materials with synthesis-dependent structural and optical properties [22]. Remarkably, direct HPHT synthesis from adamantane precursors has been shown to produce nanodiamonds hosting SiV centers with nearly Fourier-transform-limited optical transitions, demonstrating that excellent optical quality can also be achieved through HPHT-based nanodiamond synthesis routes [21]. While SiV-containing NDs produced by at HPHT synthesis have demonstrated promising optical properties, the synthesis method based on metal-free growth under extreme conditions requires initial complex chemical mixtures and yields limited quantities. In contrast, microwave-assisted CVD represents a potentially scalable route for producing high-purity NDs, while maintaining a high degree of control over impurity incorporation and crystal growth conditions [17]. This substrate- and seed-free process simplifies particle recovery, reduces contamination, and allows fine control of dopant incorporation. More specifically, CVD offers a versatile approach for the synthesis of high-purity, high-quality SiV-containing diamond particles. In particular, Zhang et al. [23] used salt-assisted air-oxidized NDs as CVD seeds to grow high-quality diamond microparticles containing SiV centers, which exhibited enhanced photoluminescence intensity, narrow optical linewidths, and reduced inhomogeneous broadening. On the other hand, De Feudis et al. [17] developed a seedless CVD growth strategy for the production of free-standing nanodiamonds with high crystalline quality, facile recovery, reduced contamination, and fine control over dopant incorporation. A more recent alternative CVD-based approach has also demonstrated high-yield production of SiV-containing diamond nanoparticles with controlled optical properties, highlighting the growing maturity of CVD synthesis routes for quantum-grade nanodiamond materials [14]. CVD diamond particles have recently been employed in diamond anvil cells, where pressure-dependent photoluminescence was measured up to 180 GPa [9,17]. These experiments reveal pressure-induced spectral shifts consistent with theoretical predictions, positioning CVD SiV-NDs as promising quantum sensors capable of operating in regimes challenging for other color centers. However, low-temperature optical studies of SiV nanodiamonds still reveal substantial line broadening that prevents full resolution of the fine structure and limits quantum performance. This broadening is commonly associated with residual strain, local lattice distortions, and structural defects, all of which can significantly affect the spectral position and linewidth of SiV optical transitions [24,25]. Post-growth annealing is therefore an attractive route to reduce structural disorder and improve spectral homogeneity. However, in nanoscale diamond systems the accessible annealing temperature range remains strongly constrained by graphitization. Previous studies demonstrated that high-pressure high-temperature (HPHT) annealing treatment can promote defect recovery, improve crystal quality, activate vacancy-related centers, and suppress graphitization compared with ambient-pressure annealing conditions. Several representative examples are summarized below. HPHT annealing treatments have been successfully applied to natural, HPHT-grown, CVD-grown, and implanted diamonds, where pressure plays a key role in maintaining diamond stability while enabling defect migration and lattice reorganization. In particular, HPHT annealing has been shown to improve the optical quality of diamond by reducing defect-related strain and facilitating the evolution of vacancy-impurity complexes [26], while also reducing inhomogeneous broadening in SnV-containing diamond films [27]. More recently, HPHT post-treatment of implanted nitrogen-vacancy (NV) centers was reported to increase color center density, improve spin coherence, and reduce internal stress in diamond quantum sensors operating up to 240 GPa [28]. For particulate systems, Nunn et al. showed that vacuum annealing of micrometre-sized diamond particles containing NV centers improved optical contrast and reduced spectral broadening, although the study was limited to ambient pressure, NV defects, and relatively large particles less prone to graphitization [29]. On the other hand, Li et al. [30], demonstrated that the onset of graphitization in micrometric diamond powders strongly depends on pressure and stress state. In particular, quasi-hydrostatic conditions can substantially delay graphite formation compared with non-hydrostatic compression, highlighting the importance of the thermodynamic environment during annealing treatments. However, these investigations were performed on micrometric diamond powders and did not address the phase stability of quantum nanodiamonds containing color centers. Recent studies have demonstrated that HPHT processing can provide an effective route to improve the quality of color-center-containing NDs while suppressing graphitization. In particular, Bao et al. [31] reported the production of quantum-grade NV NDs through a pressure- and temperature-assisted process at 7 GPa and 1700 °C, in which NaCl was directly mixed with the nanodiamond powder and acted both as a sintering inhibitor and, upon melting, as a pressure-transmitting medium that contributed to maintaining diamond stability at elevated temperatures. These results highlight the potential of HPHT treatments for optimizing quantum ND materials. However, despite these advances, direct experimental determination of graphitization boundaries in ND systems remains limited. *In situ* diffraction studies capable of following structural evolution under controlled pressure–temperature conditions are particularly scarce. The present work addresses this aspect by combining synchrotron energy-dispersive X-ray diffraction with HPHT processing and applying the resulting pressure–temperature window to the annealing of SiV-containing NDs. In this work, we use a Paris–Edinburgh (PE) press coupled with *in situ* synchrotron X-ray diffraction to determine the graphitization threshold of nanodiamonds under HPHT conditions and to define a safe pressure–temperature processing window. Calibration experiments were first carried out on type-Ib nanodiamond powders, followed by targeted annealing treatments of CVD-grown SiV nanodiamonds.

* Corresponding author. Institut de Recherche de Chimie Paris, Chimie ParisTech, UMR CNRS 8247, PSL Research University, Paris, 75005, France. CY Cergy Paris Université, Cergy-Pontoise, 95031, France
*E-mail address:* mary.de-feudis@cyu.fr (M. De Feudis).

## 2. Experimental details

### 2.1. Nanodiamond materials

Two different nanodiamond systems were used in this study. The calibration experiments were performed using commercially available type-Ib HPHT-derived nanodiamonds obtained by milling bulk HPHT diamond (Element Six). Dynamic light scattering (DLS) measurements performed after brief ultrasonication revealed a dominant particle population centred at approximately 129 nm in the number-weighted distribution. The corresponding volume-weighted distribution indicates the presence of a limited fraction of larger agglomerates. Both representations are provided in the Supplementary Material (Section S1, Fig. S1) and provide complementary information on the initial suspensions. Additional XPS and representative SEM analyses confirmed a predominantly diamond-like surface chemistry together with a nanoscale morphology consistent with the DLS measurements. Both XPS and SEM characterizations are reported in Supplementary Material (Section S1, Fig. S2–S3, Table S1). These (Ib-NDs) powders were selected for calibration experiments because they were available in sufficiently large quantities (tens of milligrams), allowing reproducible capsule loading and reliable *in situ* synchrotron diffraction measurements.

CVD-grown SiV nanodiamonds (CVD-NDs) were used for the targeted annealing experiments and subsequent optical characterization. These particles were synthesized by CVD following the procedure described in Ref. [17]. DLS measurements performed after brief ultrasonication revealed a dominant particle population centred at approximately 147 nm in the number-weighted distribution. The corresponding volume-weighted distribution indicates the presence of a fraction of larger agglomerates, particularly for the CVD-grown nanodiamonds. Both representations are reported in the Supplementary Material (Section S1, Fig. S1) and should be considered complementary descriptors of the initial suspensions. XPS analyses showed a surface chemistry broadly comparable to that of the Ib-NDs, with diamond-like carbon as the dominant component together with oxygen-containing surface functionalities. XPS and representative SEM analyses are reported in Supplementary Material (Section S1, Fig. S2–S3, Table S1). The specific CVD batches used in this study were produced in limited quantities at the laboratory scale (sub-milligram scale), therefore these samples were reserved for optimized HPHT treatments identified from the calibration experiments.

### 2.2. *As-grown SiV-containing nanodiamond particles and pre-annealing characterization*

The SiV-doped nanodiamonds used in this study were synthesized by the in-house seedless and substrate-independent CVD protocol previously reported in Ref. [17]. Silicon incorporation was achieved by introducing a solid silicon source within the reactor chamber during deposition, allowing the formation of luminescent silicon-vacancy centers within the diamond lattice. The addition of a controlled amount of $N_2$ gas to the $H_4/CH_4$ mixture during growth promotes the activation of the negative charge state SiV as an electron donor do to induce an intense photoluminescence (PL) signal at 738 nm. The resulting nanodiamonds exhibit high crystallinity and chemical purity, as confirmed by Raman spectroscopy showing a diamond Raman peak near 1330 $cm^{-1}$ with a full width half maximum (FWHM) of about 10 $cm^{-1}$, with no significant evidence of graphite or amorphous carbon [17]. No post-synthesis chemical treatments were necessary due to the clean growth conditions. These CVD-grown nanodiamonds containing $SiV^-$ centers (briefly labelled as *SiV* centers) constitute the main focus of this work and were used for all low-temperature optical measurements and targeted HPHT annealing. For the calibration experiments, however, gram-scale quantities of ND powder were required to reproducibly fill the Paris–Edinburgh press assembly. Therefore, commercially available Ib-NDs were used for the calibration runs, whereas the limited-quantity CVD-grown SiV nanodiamonds were reserved for the selected HPHT annealing experiments. For this work, new batches of nanodiamonds were synthesized to investigate their performance at low temperatures with the aim resolving the SiV fine structure. Fig. 1a shows the photoluminescence spectrum of a small cluster of nanodiamonds containing SiV centers: the zero-phonon line emission of the $SiV^-$ centers at 738 nm is so intense that it reduces the apparent intensity of the diamond peak at 504.7 nm (enlargement in the inset). A small shoulder visible at ambient conditions on the right-hand side of the line is attributed to the phonon sideband.

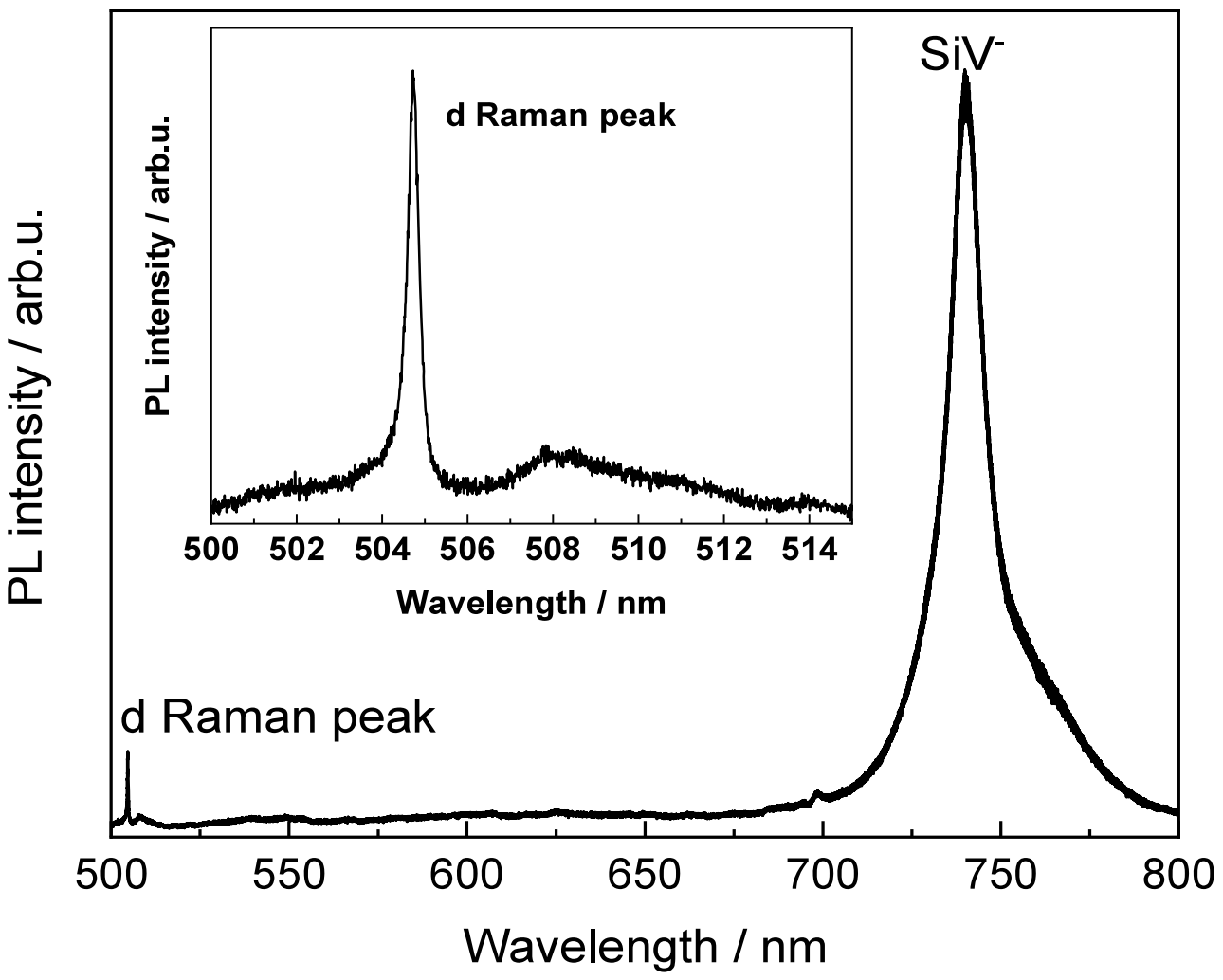

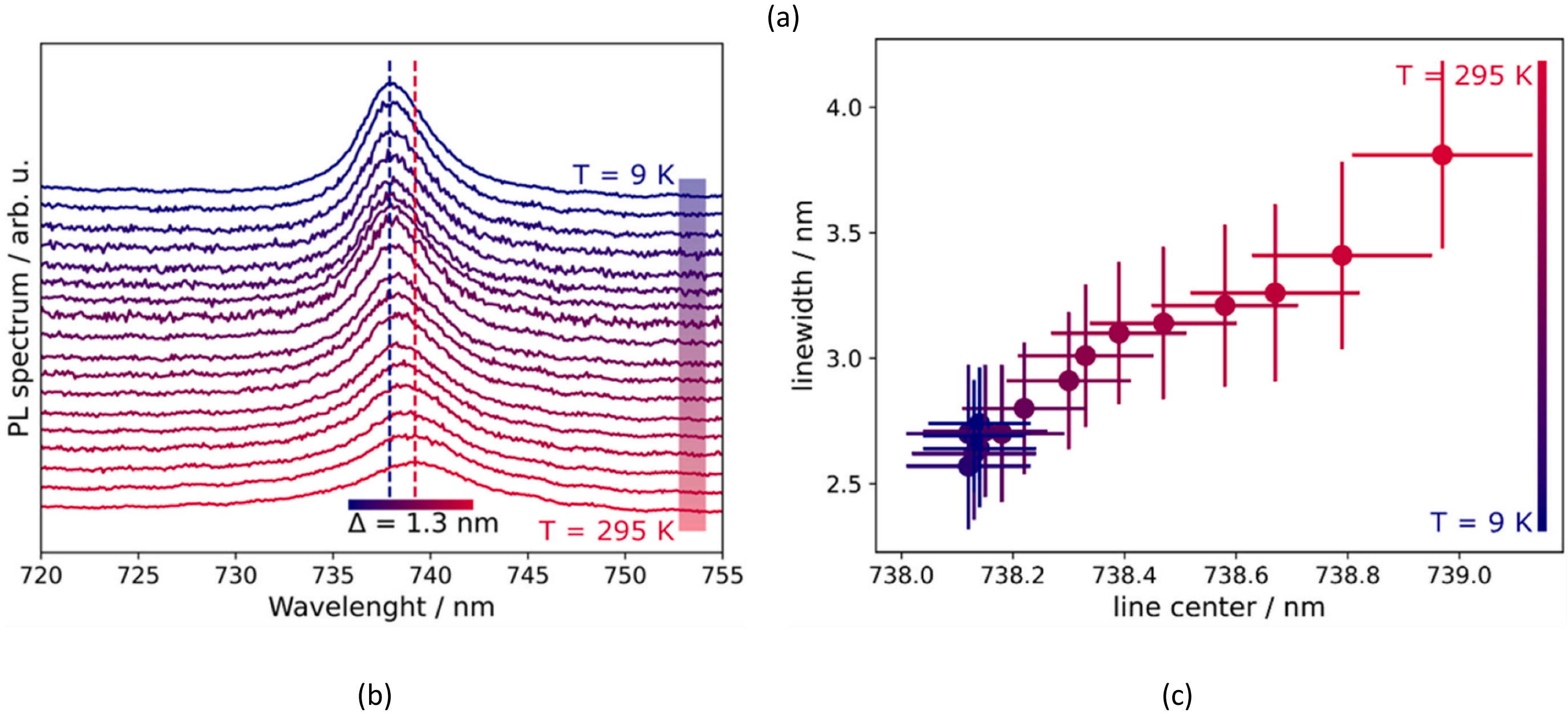

**Fig. 1.** (a) PL spectrum registered at room temperature on a Si-doped as-grown ND cluster showing the highly emissive SiV centers (excitation at 473 nm). Inset: magnification of the d Raman peak (range 500-515 nm). The close small band at around 508 nm (1450 $cm^{-1}$) can be ascribed to vibrations of trans-polyacetylene molecules, typical of the CVD diamond synthesis technique [32]. (b) PL spectra of SiV centers registered on as-grown NDs showing the evolution of the ZPL peak as a function of the temperature down to 9 K (step ≈ 15 K, excitation at 532 nm); (c) corresponding correlation between line center and linewidth, displaying an approximately linear trend.

The optical quality of the as-grown SiV NDs was evaluated through PL spectroscopy at temperatures in the range of 295 K down to 9 K. As shown in Fig. 1b, the low-temperature PL spectrum still displays broadened emission lines, which prevent the resolution of the characteristic four-line fine structure of the silicon-vacancy center. This broadening arises from synthesis-induced local strain fields and an inhomogeneous distribution of defects within the nanodiamond lattice, which together limit the potential for high-fidelity quantum manipulation. The broadening of the SiV emission provides the rationale for investigating post-growth annealing treatments in the present work. Remarkably, a progressive blue shift of the ZPL is observed upon cooling as shown in Fig. 1c, consistent with the known sensitivity of the color centers' optical properties to the surrounding environment, accompanied by a pronounced but partial narrowing of the FWHM as the temperature is lowered [8].

### 2.3. High-pressure high-temperature annealing experiments

In order to mitigate local strain in diamond, post-synthesis annealing treatments are commonly employed. Indeed, diamonds are known to show crystalline quality improvement when annealed under vacuum or at ambient pressure. The supplied thermal energy allows relaxation of local strain associated with structural defects and lattice imperfections generated during nanodiamond growth. Such HPHT-assisted defect recovery processes have previously been discussed in diamond annealing studies, where vacancy migration and lattice reorganization contribute to improved crystal quality and optical performance [26,33]. The annealing treatment also increases the probability of coupling between vacancies and impurities hence increasing the density of color centers. This leads the enhancement of their emission at the ZPL. At ambient pressure, however, the annealing temperature is intrinsically limited by the onset of the diamond-to-graphite phase transition, which typically occurs at $\approx 700\,^{\circ}$C in air for diamond films [34]. Depending on the surrounding atmosphere, for instance under vacuum or inert gases, the temperature threshold may vary between 1100 and 1300 °C [35], but it cannot be exceeded without risking partial graphitization. For nanodiamond systems, the situation is even more critical. Due to their high surface-to-volume ratio, surface graphitization may start already at $\approx$ 900 °C under ultra-high vacuum [36], resulting in structural degradation and optical quenching of color centers. To overcome these limitations, we employed HPHT annealing using a large-volume PE press. The application of hydrostatic pressures, even at a few of gigapascals, stabilizes the diamond phase well above the graphitization temperature at ambient pressure according to the carbon phase diagram [37], allowing a better crystal lattice relaxation and thermal activation of vacancy–impurity reorganization at higher temperatures without any undesired graphite formation. This high-pressure environment also ensures a more homogeneous heat distribution and prevents local thermal runaway effects, which are particularly detrimental in nanoscale diamond powder.

HPHT experiments were performed at the PSICHE beamline of the SOLEIL synchrotron using the Ultrafast Tomography Paris–Edinburgh Cell (UToPEC). In situ monitoring combined energy-dispersive X-ray diffraction (ED-XRD), X-ray tomography (XCT), and radiography to follow the structural and phase evolution of the samples during compression and heating. Additional details on the beamline instrumentation and acquisition geometry are provided in the Supplementary Material (S2). The UToPEC setup operates up to $\sim$10 GPa and above 2000 K, and can also be used in off-beam mode for calibrated annealing treatments [38]. The press is equipped with an on-board automatic compressor and the electrical cables and coolant hoses, which are routed through the rotary hole of the press (see Fig. 2a). This configuration enables precise control of pressure, heating and cooling rates [39]. To identify the pressure–temperature conditions suitable for HPHT annealing of SiV-containing nanodiamonds, a systematic study was performed over the 2–4 GPa pressure range and temperatures between 1000 and 2000 K. The recovered samples were subsequently characterized by photoluminescence spectroscopy. For low-pressure experiments (2 GPa), we used the larger PE press assemblies, hereafter referred to as 10 mm assemblies, which provide a larger sample chamber. Higher pressures (such as 4 GPa) were achieved using 5 mm assemblies. Calibration runs were first performed for both assemblies to assess the pressure and temperature conditions, with the dual aim of: (i) directly investigating, for the first time, the transformations occurring in our nanoparticle samples, and (ii) establishing calibration curves for subsequent off-line experiments. To ensure meaningful calibration, particular attention was devoted to the sample configuration inside the pressure assembly. One could argue that the current setup employed is non-hydrostatic and could increase the lattice strain in nanodiamond and favour sintering upon pressurizing. In order to address this issue, the nanodiamond were dispersed in a salt powder of NaCl preventing from aggregation of nanodiamonds and taking advantage of its low melting temperatures under pressure ($\sim$1000 K, [40]) to ensure hydrostatic conditions. The 10 mm assembly (Fig. 2b) was composed of standard components, which are here schematically represented and listed (Fig. 2c): (i) a hexagonal boron nitride (hBN) cylindrical capsule used to host the sample and to act as the primary pressure-transmitting medium; (ii) a graphitic cylinder and caps acting as the furnace; and (iii) a boron–epoxy gasket housing the graphite furnace. Molybdenum disks and magnesium oxide–steel rings were used to complete the assembly above and below the gasket aperture, ensuring optimal electrical contact for high-temperature generation. Finally, the full assembly was fitted into a PEEK ring to prevent extrusion during compression (visible in Fig. 2b). Accordingly, the sample had to be strategically packed into the hBN capsule (internal diameter 2.4 mm, height 2.5 mm). In addition to the 10 mm configuration, a 5 mm assembly was also available, employing the same component design but with reduced capsule dimensions (internal diameter 1.5 mm, height 2.0 mm). This more compact assembly enables the system to sustain higher pressures and, consequently, to reach higher temperatures.

In order to determine the assembly response to load and power, we performed a calibration run for which we designed a sandwich-type sample configuration, as shown in Fig. 2d. The capsule was filled manually with three sequential layers so organised: milled NaCl powder (1/3 of the total volume) mixed with around 10 platinum (Pt) spheres on the bottom, nanodiamonds (1/3) in the middle, NaCl powder (1/3) mixed with around 10 platinum (Pt) spheres on the top. This NaCL(Pt)-ND-NaCl(Pt) configuration allowed the NDs to be embedded in an easily manageable and stable medium, while NaCl played a dual role: handling medium and a quasi-hydrostatic pressure-temperature transmitting medium during sintering with a well-characterized thermal equation of state and melting curve [40] for which X-ray diffraction allowed precise temperature calibration during synchrotron experiments. The Pt spheres inclusion allowed to constrain the furnace efficiency at temperature above 2000 K by means of X-ray imaging observation of their melting.

Based on these calibration experiments, selected pressure–temperature conditions were subsequently applied to CVD-grown SiV nanodiamonds. After annealing, the recovered sample consisted of a solid NaCl block containing dispersed NDs. The recovery procedure involved dissolution of the NaCl matrix in distilled water assisted by thermal treatment (95 °C) and ultrasonic cleaning, followed by repeated centrifugation and washing cycles to separate the nanodiamond fraction from the dissolved salt. The recovered particles were subsequently dispersed and prepared for SEM, EDX and optical characterization. Full details of the recovery protocol are provided in Supplementary Material Section S4.

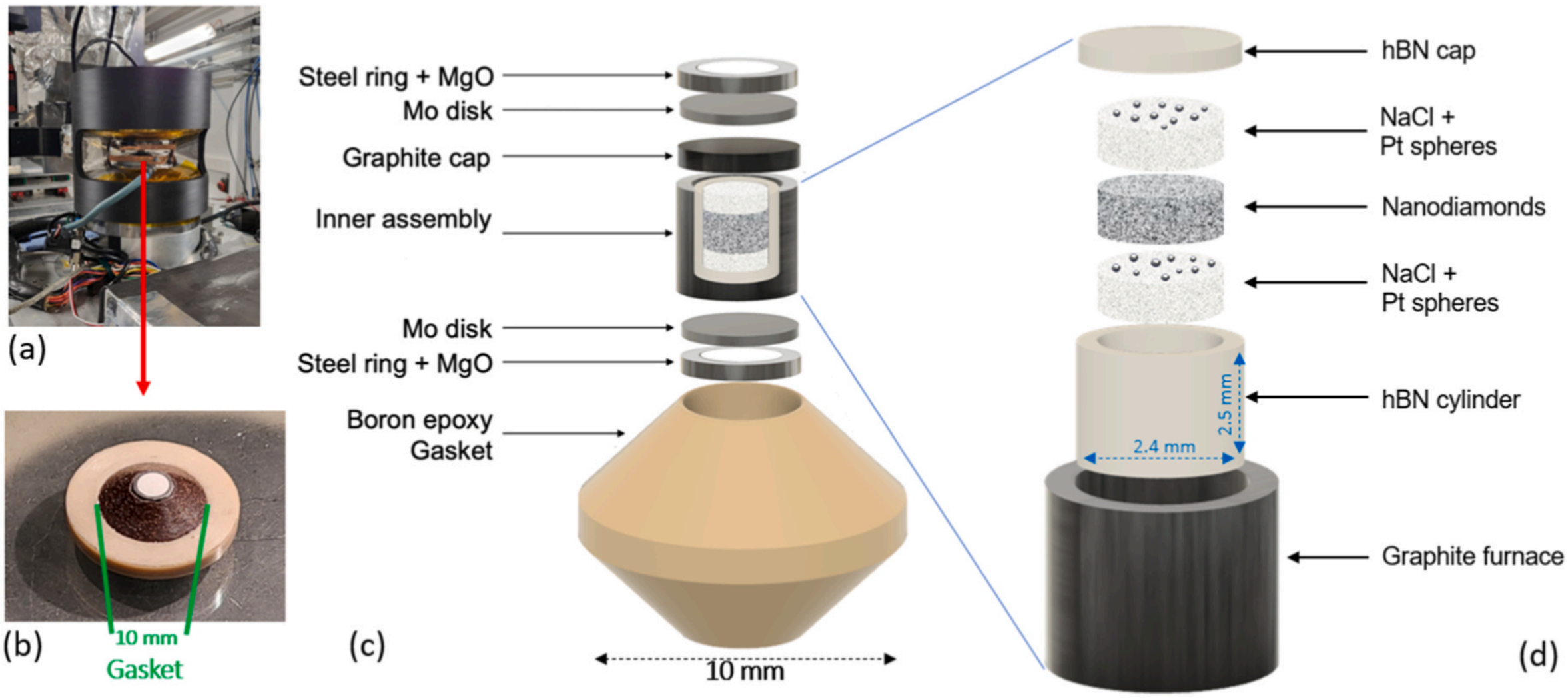

**Fig. 2.** (a) PE press installed at the PSICHE beamline of the SOLEIL synchrotron. (b) Photograph of the final 10 mm assembly. (c) Exploded view of the assembly components inside the gasket: the sample is enclosed in a hBN capsule, which is placed inside a graphite furnace, while three disks (graphite, molybdenum, and a magnesium oxide–steel ring) are positioned above and below the capsule to ensure uniform pressure and temperature conditions. (d) Sandwich-type sample configuration inside the hBN capsule, from bottom to top: a layer of milled NaCl salt grains and Pt spheres, a layer of ND particles, and a final layer of milled NaCl salt grains and Pt spheres.

### 2.4. *In situ experimental procedure at the SOLEIL synchrotron*

The standard operation protocol involved placing the final sample assembly between the tungsten carbide (WC) anvils of the press and increasing the pressure to the desired value. After pressurization at room temperature, the temperature was progressively increased at constant load to preserve the electrical contacts and ensure stable heating conditions. The applied mechanical load and the electrical power supplied to the graphite furnace were directly controlled to modulate pressure and temperature, respectively.

Pressure and temperature were monitored *in situ* by combining energy-dispersive X-ray diffraction, X-ray tomography, and radiography. In contrast to conventional calibrations relying on external thermocouples, the internal temperature of the assembly was directly determined from the evolution of crystallographic planes measured by XRD. In particular, NaCl X-ray diffraction was used to track pressure and temperature up to its melting point, while the onset of nanodiamond graphitization, as monitored by XRD, provided an additional high-temperature reference. The presence of Pt spheres ensured a high temperature reference point corresponding to possible melting. This approach enabled a more accurate determination of the internal temperature experienced by the sample compared that measured with a thermocouple, which would have had to be necessarily located away from the sample.

It is worth noting that performing the same calibration experiment on two assemblies with different sizes allowed us to study the morphological and structural changes in the NDs sample at two different pressures, namely, 2 and 4 GPa, for the 10 mm and 5 mm configurations respectively, as the temperature increased. This paves the way for the accurate construction of a graphitization diagram of diamond nanoparticles, whereas to date there have been clear studies on massive diamond [37].

### 2.5. *Synchrotron instrumentation and beamline setup*

The synchrotron instrumentation and beamline setup are described in detail in Supplementary Material (S2).

### 2.6. *Optical setups*

Room-temperature Raman and PL measurements were carried out using a Renishaw InVia Raman/PL system equipped with 532 nm and 473 nm laser excitation lines. The excitation beam was focused onto the sample using a 100× objective, allowing localized spectroscopic analysis.

Cryogenic spectroscopic characterizations were performed using home-built confocal microscopy systems. Excitation was provided by 532 nm and 594 nm laser lines, focused with a 100× objective.

The nanodiamonds dispersed in ethanol were drop-cast onto silicon substrates and allowed to air dry prior to optical measurements.

### 2.7. *Electron microscopy and elemental analysis*

Optical images of the calibration sample “CalibND10” after HPHT annealing in the PE press were acquired using a Keyence VHX-5000 digital confocal microscope, after extraction from the high-pressure assembly.

Dynamic light scattering (DLS) measurements were performed using a Malvern Zetasizer Nano ZS instrument. The analyses were carried out on ethanol suspensions after brief ultrasonication of the nanodiamond samples using a Hielscher UP100H ultrasonic processor. Particle-size distributions were obtained from number-weighted distributions.

X-ray photoelectron spectroscopy (XPS) measurements were performed to investigate the surface chemistry of the two nanodiamond systems. The analyses were carried out using a PHI Genesis instrument (Physical Electronics, Chanhassen, MN, USA) equipped with a monochromatic Al Kα X-ray source (hν = 1486.6 eV). High-resolution C1s spectra were acquired to evaluate the carbon bonding environment and surface functional groups present on the nanodiamond surfaces. Surface charging effects were compensated using the instrument's dual-mode charge neutralisation system combining low-energy electrons and ions.

Scanning electron microscopy (SEM) and energy-dispersive X-ray spectroscopy (EDX) analyses were performed using a ZEISS Gemini SEM

360 instrument equipped with an Oxford Instruments Ultim Max 100 EDX detector. These analyses were conducted during the development of the nanodiamond recovery protocol (centrifugation, thermal treatments, and ultrasonics cleaning steps) in order to assess morphology and elemental composition at different stages of the purification process. The full ND recovery protocol is reported in Supplementary Material (S4).

The nanodiamonds dispersed in ethanol were drop-cast onto silicon substrates and allowed to air dry prior to microscopy and spectroscopy analyses.

## 3. Results and discussion

### 3.1. *In situ tomographic monitoring of the NaCl matrix*

In situ X-ray tomography was used to monitor the positioning and integrity of the assembly during pressurization and to follow the structural evolution of the NaCl matrix during the initial stages of heating. Representative tomographic slices acquired at increasing temperature and at a fixed pressure of approximately 2 GPa for the sample in the 10 mm configuration are shown in Fig. 3.

At room temperature and ambient pressure, the assembly exhibits well-defined interfaces between the NaCl matrix and the nanodiamond layer, with bright platinum spheres clearly visible (highlighted by yellow cercles) and no evidence of deformation or collapse of the graphite furnace and hBN capsule (Fig. 3a). After pressurization to 2 GPa at room temperature, a progressive reduction of porosity is observed within the NaCl matrix, while the overall geometry of the assembly remains preserved, indicating efficient compaction under pressure (Fig. 3b). Upon heating to 925 K, the NaCl phase becomes more homogeneous, accompanied by subtle textural changes consistent with grain coarsening and recrystallization (Fig. 3c). At higher temperature (T = 1425 K), a marked loss of internal texture and contrast is observed within the salt matrix (Fig. 3d), indicating the onset of NaCl melting. This interpretation is independently confirmed by the disappearance of NaCl peaks in the simultaneously acquired XRD patterns (see next section).

At this stage, the nanodiamonds become dispersed within the molten salt, while the platinum spheres remain clearly detectable, although slightly displaced, suggesting buoyancy or sinking effects in the liquid medium. Throughout the tomographic sequence, the platinum spheres remain intact and well defined, providing an additional upper bound for the maximum temperature reached during the experiment. Overall, the tomographic observations provide direct spatially resolved evidence of the thermal evolution of the pressure-transmitting medium. They also demonstrate the mechanical integrity of the assembly under extreme pressure–temperature conditions and identify NaCl melting as an internal temperature marker for the subsequent calibration of the experimental setup.

Complementary tomographic reconstructions of the 5 mm assembly, illustrating sample positioning and compaction under pressure at room temperature, are provided in the Supplementary Material, S3 and Fig. S4.

### 3.2. *In situ X-ray diffraction analyses: PE press calibration and study of the nanodiamonds structural evolution*

*In situ* ED-XRD was employed to constrain pressure and temperature conditions throughout the HPHT experiments by exploiting multiple internal crystallographic markers. The hBN capsule surrounding the sample provided a continuous and reliable pressure reference over the entire temperature range explored. Its lattice parameters were extracted from X-ray diffraction patterns acquired at each temperature step and converted into pressure values using an established equation of state, allowing pressure to be determinated independently of the applied mechanical load. This approach compensates for load-dependent effects arising from cell deformation and enables direct tracking of the pressure experienced by the sample during heating. The use of hBN as an internal pressure calibrant is particularly advantageous under HPHT conditions as it remains crystalline and chemically inert over the investigated pressure–temperature range and provides stable and well-resolved X-ray diffraction peaks suitable for accurate equation-of-state (EOS)-based analysis [41]. The full ED-XRD datasets for the 10 mm and 5 mm assemblies are reported in the Supplementary Material S3, Fig. S5(a) and S5(b).

In parallel, X-ray diffraction from the NaCl pressure-transmitting medium provided a direct internal temperature marker up to its melting point. Upon heating at constant pressure, NaCl diffraction peaks progressively shift toward lower energies and decrease in intensity, reflecting thermal expansion and grain coarsening. The complete disappearance of X-ray diffraction peaks marks the onset of salt melting, Fig. S5(a), Supplementary Material. In the 10 mm assembly, NaCl X-ray diffraction remains detectable up to approximately 1425 K at 2 GPa, in excellent agreement with the melting conditions reported by Akella et al. [40] under comparable pressure conditions and shown in Fig. 4). For the 5 mm assembly, NaCl melting occurs at higher temperature 1770 K at 4 GPa (Fig. S5(b) available in Supplementary Material), consistent with the higher-pressure conditions achieved in this configuration and very close to published results (almost 1400 °C) [16]. The systematic shift and eventual loss of NaCl X-ray diffraction thus provide a robust and reproducible criterion for identifying the melting temperature of the pressure-transmitting medium and for calibrating the internal temperature of the assembly independently of external thermocouple readings.

For completeness, Fig. 4 shows a comparison of the pressure–temperature calibration curves obtained for the two assemblies using the hBN equation of state adopted in this work [41] with those derived from two other widely accepted EOS references ([42], [43]). A good overall convergence is observed among the different calibrations, indicating that the pressure determination is only weakly dependent on the specific EOS parameterization within the explored P–T range. Furthermore, the NaCl melting curve reported by Akella et al. [40] intersects the experimentally determined melting points obtained in the present study, providing an additional and independent validation of the pressure–temperature calibration.

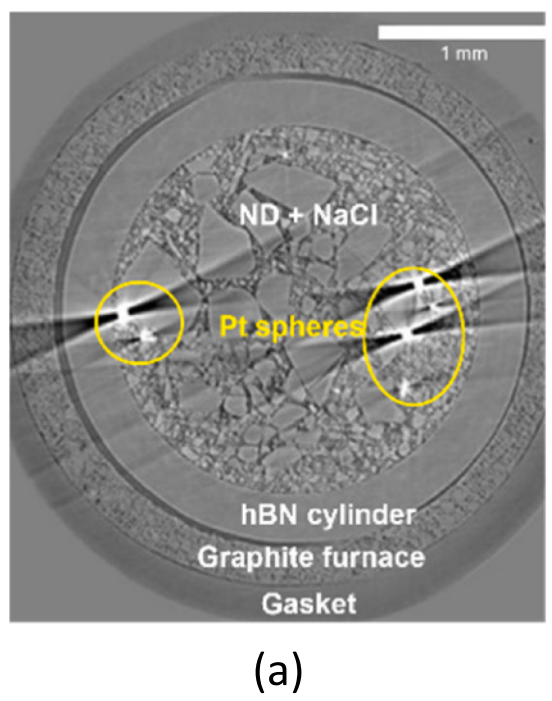

(a)

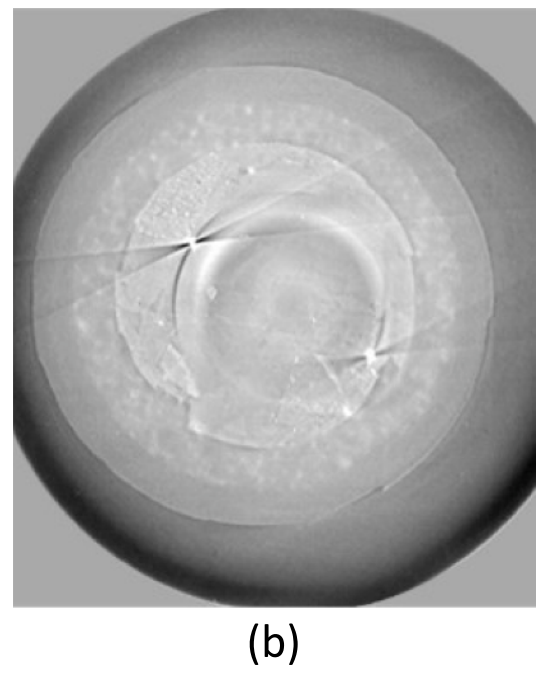

(b)

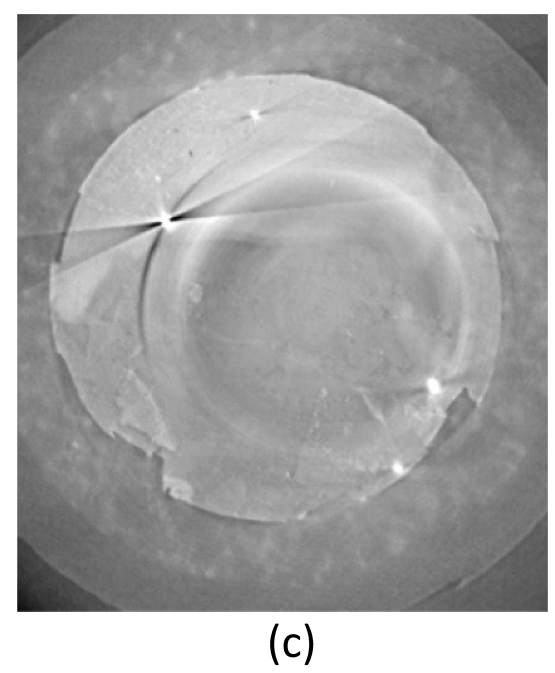

(c)

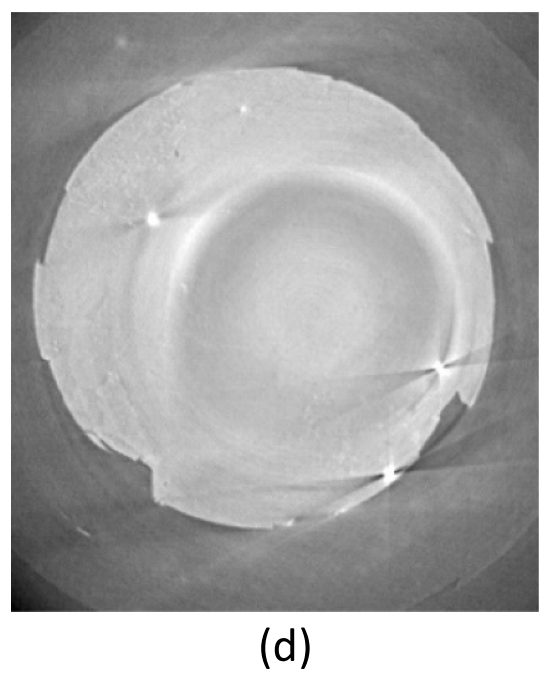

(d)

**Fig. 3.** Tomographic sequence of the 10 mm assembly recorded during HPHT annealing using a PE press under different pressure and temperature conditions: (a) at room temperature and ambient pressure, the sample is correctly positioned with respect to the X-ray beam, and the various components of the assembly are clearly visible; (b) at 2 GPa and room temperature, the assembly components remain preserved while salt (NaCl) densification initiates; (c) at 2 GPa and 925 K, enhanced compaction leads to the onset of salt recrystallization; (d) at 2 GPa and 1425 K, further heating results in NaCl melting (as confirmed from XRD analysis).

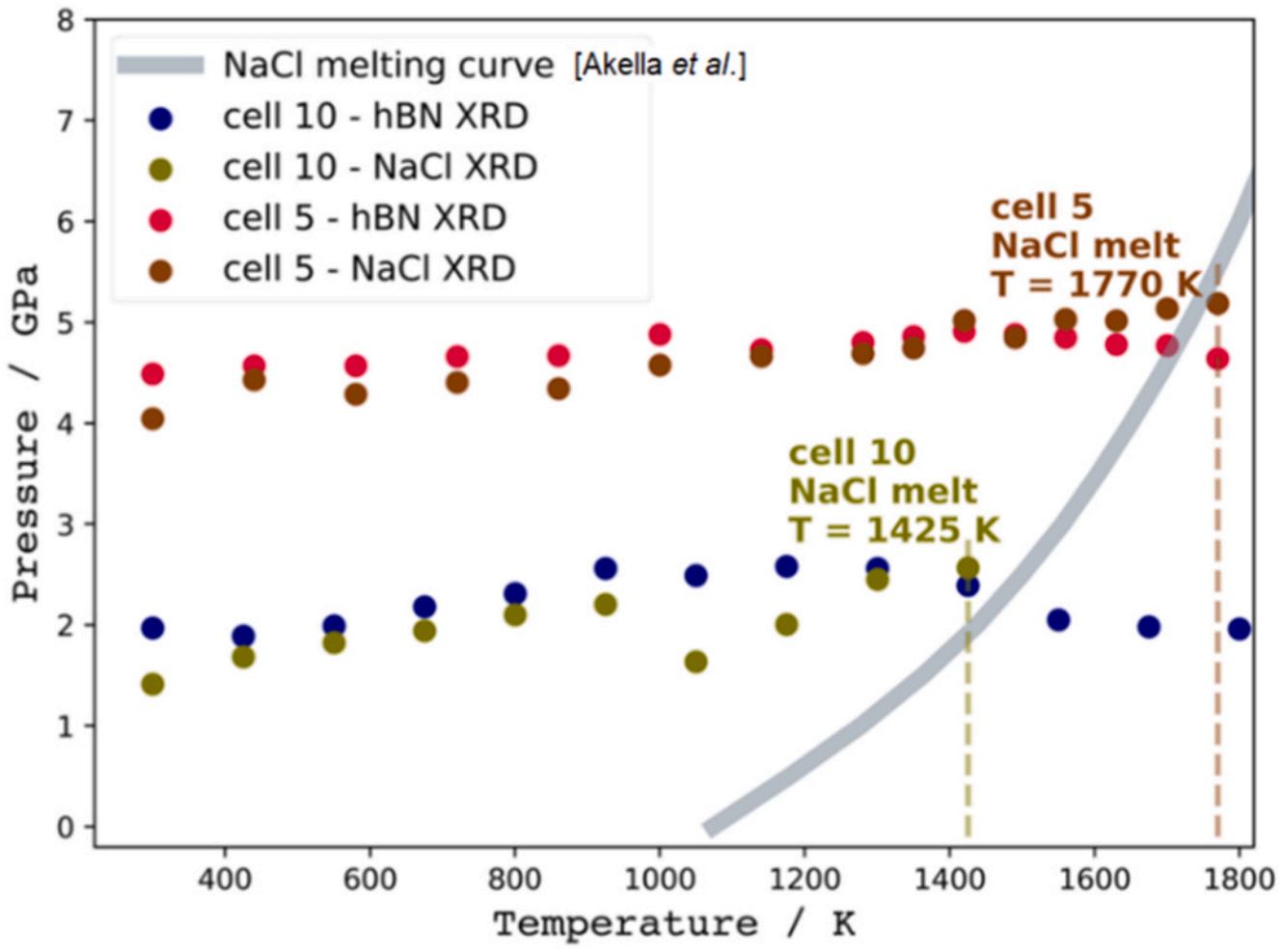

**Fig. 4.** Pressure–temperature calibration of the 10 mm and 5 mm PE press assemblies. The pressure data were obtained from experimental XRD peaks using the equations of state (EOS) for hBN and NaCl, respectively. The grey curve correspond to the NaCl melting curve reported by Akella et al. [40] and intersects the experimentally determined melting points (T = 1425 K @ 2 GPa and T = 1770 K @ 4 GPa), confirming the reliability of the pressure–temperature calibration.

The evolution of the internal temperature with applied electrical power was quantified up to the NaCl melting point for both assemblies. The equations are reported in the Supplementary Material, S3.

*In situ* ED-XRD was further employed to investigate the structural evolution of the nanodiamonds (with characteristic sizes of 100 – 200 nm) as a function of temperature under controlled pressure conditions (2 – 4 GPa). ED-XRD patterns were collected at a fixed diffraction angle and analysed using calibrated energy spectra to determine d-spacings and identify crystalline phases. Phase identification was performed by comparison with reference diffraction patterns of diamond, graphite, NaCl and hBN. The evolution of the corresponding diffraction peaks during heating allowed *in situ* monitoring of structural changes, including NaCl melting and the onset of nanodiamond graphitization. Building upon the pressure–temperature calibration established from hBN and NaCl X-ray diffraction, the experiments were extended to higher temperatures in order to directly identify the graphitization boundary of nanodiamonds under controlled pressure conditions. The onset of graphitization was unambiguously identified by XRD through the appearance of characteristic graphite X-ray diffraction peaks. To facilitate visualization of the structural evolution, the diffraction features associated with hBN, NaCl, diamond and graphite are displayed together in Fig. 5a and b, while Fig. 5c and d shows an enlarged view of the spectral region containing the graphite reflection.

In particular, Fig. 5a shows the XRD patterns collected from nanodiamonds during heating at three key temperatures under 2 GPa: 300 K (reference), where the diamond crystal planes ((111), (220), (311)) are dominant, confirming preservation of the crystalline diamond structure; 1425 K, corresponding to NaCl melting, as indicated by the disappearance of the salt diffraction peaks ((111), (200), (220), (222)), while diamond reflections remain clearly detectable; and 1800 K, where the onset of graphitization is evidenced by the emergence of the graphite (002) reflection, indicating significant transformation of diamond into graphite. A similar analysis for experiments at 4 GPa, including the corresponding diamond, NaCl, and graphite crystallographic planes, indicates that diamond reflections ((111), (220), (311)) persist to higher temperatures, with the diamond–graphite phase transition occurring at 2120 K (Fig. 5b). The simultaneous presence of diamond and graphite XRD peaks is also observed at these temperatures, in line with the onset of diamond-graphite phase transition and with the primary objective of this work, namely to define a safe annealing temperature that prevents graphitization, while a detailed investigation of the kinetics of the diamond-to-graphite transformation will be addressed in follow-up experiments. Tracking the temperature evolution of the XRD peaks in Fig. 5c and d, a monotonic shift of the central peak attributed to hBN is first observed. Additionally, the NaCl peaks disappear as the temperature crosses its melting point. Finally, the appearance of the graphite peak on the low-angle side of the hBN peak allows determination of the graphitization onset temperatures for both assemblies, corresponding to the two pressure conditions. The progressive evolution of the diffraction features reveals the structural stability window of nanodiamonds under HPHT conditions and establishes the temperature threshold for graphitization within the investigated pressure range. The experimentally determined graphitization onsets provide direct experimental reference points for the HPHT processing of nanodiamonds. To place these results in context, we compare our measurements with the phase diagram reported by Balmer et al. [37], onto which our data are superimposed in Fig. 6. The light-green region denotes the P–T range explored in this work, where diamond remained structurally stable, while the dark-green symbols indicate the experimentally determined onset of graphitization at 2 GPa and 4 GPa. Specifically, graphitization was observed at 1800 K under 2 GPa and 2120 K under 4 GPa, marking the diamond-to-graphite transition under our experimental conditions. Because the experiments start from pre-existing diamond nanoparticles, the measured graphitization onset reflects the combined influence of thermodynamic driving forces and transformation kinetics under the applied HPHT heating protocol. Additional factors may contribute to the enhanced stability of nanodiamonds under the present experimental conditions, including particle-size-related effects, kinetic limitations, the heating protocol, and the presence of NaCl. In particular, the NaCl environment may influence pressure transmission, particle confinement, and graphitization kinetics during HPHT treatment [31]. Because the present experiments were not designed to isolate these individual contributions, the respective roles of nanoscale characteristics, kinetic effects, and the surrounding medium cannot be independently assessed. The *in situ* diffraction measurements reported here nevertheless provide direct structural evidence of nanodiamond stability and graphitization thresholds under controlled pressure–temperature conditions. Overall, these results experimentally define a practical HPHT processing window for color center-containing nanodiamonds, enabling annealing treatments to be performed below the experimentally observed onset of graphitization under the investigated conditions. They also provide a calibrated framework for reliable off-beam HPHT annealing treatments that avoid detectable graphitization.

On the basis of these results, a conservative annealing temperature of 1700 K at 2 GPa was selected for subsequent HPHT treatments of nanodiamonds, ensuring operation below the graphitization threshold. It is important to note that this temperature is significantly higher than the conservative temperature typically used for nanodiamond systems, which is around 900–1000 °C under ultra-high vacuum [36]. This calibration highlights the crucial role of synchrotron-based *in situ* XRD for defining safe and reproducible operating windows in PE press experiments. More importantly, this calibrated window is directly derived from the experimentally observed graphitization onset, ensuring that annealing conditions remain within the diamond-stable region defined in Fig. 6.

Finally, this calibration enables reliable HPHT annealing experiments to be performed off-beam, without continuous synchrotron

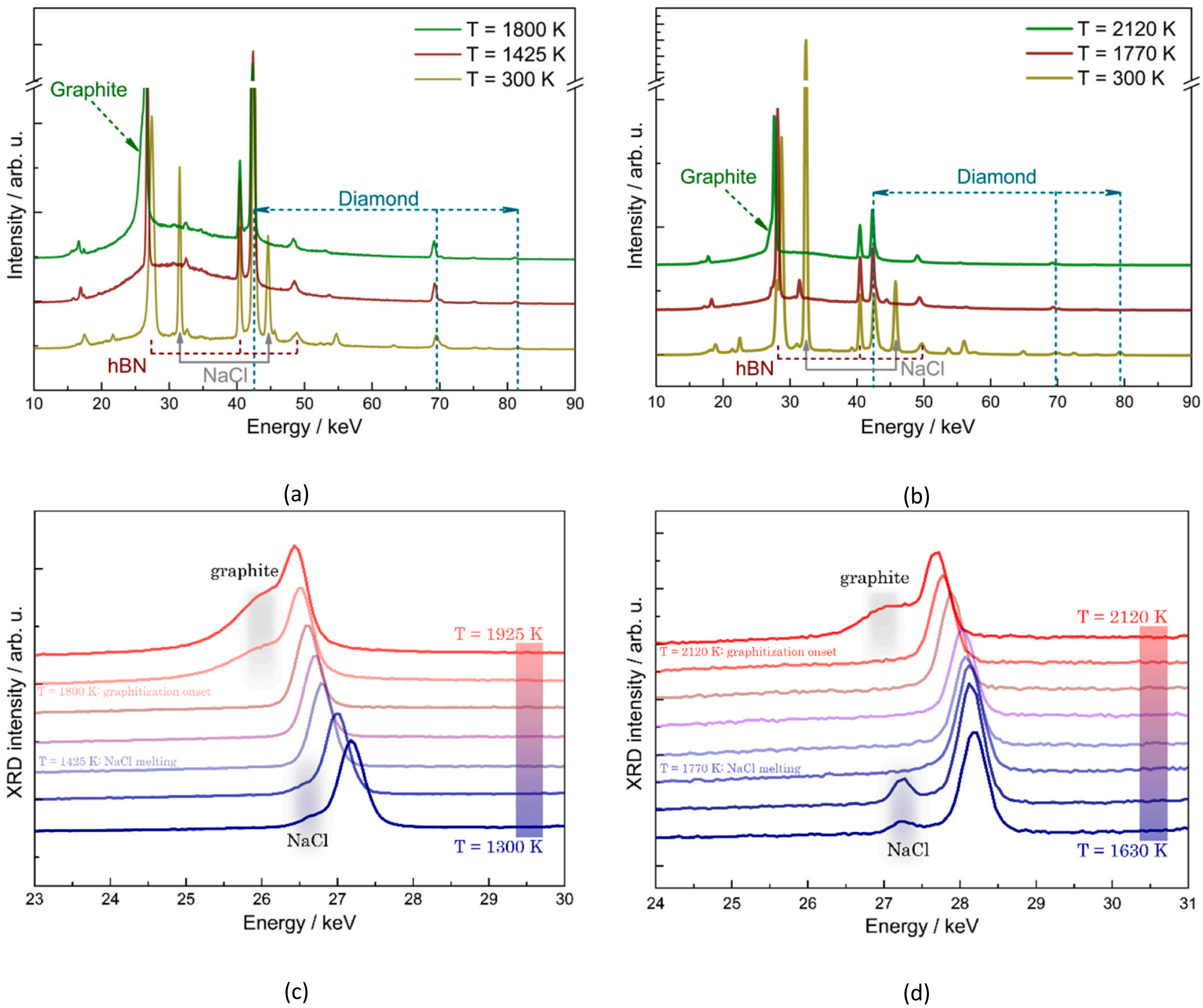

**Fig. 5.** *In situ* energy-dispersive synchrotron X-ray diffraction (ED-XRD) patterns collected from the calibration sample during HPHT annealing in the 10 mm (a) and 5 mm (b) assemblies. The main diffraction features associated with diamond, NaCl, graphite, and hBN are indicated. At room temperature, the diffraction patterns are dominated by the characteristic diamond reflections (111), (220), and (311). NaCl melting is identified by the progressive disappearance of its diffraction peaks and occurs at approximately 1425 K for the 10 mm assembly and 1770 K for the 5 mm assembly. Graphitization is detected at 1800 K and 2120 K, respectively, through the appearance of the graphite (002) reflection. Panels (c) and (d) highlight the temperature-dependent evolution of the NaCl, hBN, and graphite diffraction peaks under controlled pressure conditions (2 GPa and 4 GPa, respectively), enabling *in situ* monitoring of phase stability, salt melting, and graphitization processes.

monitoring, as applied to the samples discussed in the final section of this work.

### 3.3. In situ radiographic monitoring of Pt spheres

For completeness, the maximum temperatures reached during the calibration experiments remain below the melting temperature of platinum at the corresponding pressures. According to Errandonea [44], the expected platinum melting temperatures at the relevant pressures can be estimated using a fit to the Simon equation, yielding values of approximately 2110 K at 2 GPa and 2163 K at 4 GPa. These temperatures are significantly higher than the graphitization thresholds identified in the present experiments and therefore preclude melting of the platinum spheres under the explored conditions. Consistently, radiographic images acquired during both calibration runs show no evidence of platinum melting and are reported in the Supplementary Material, Fig. S7.

### 3.4. Ex situ recovery and characterization of calibration samples

Once the *in situ* HPHT annealing experiments were completed, pressure and temperature were returned to ambient conditions and the anvils containing the assembly were removed from the press. Following HPHT treatment, the recovered assembly was strongly deformed and highly compacted, preventing manual disassembly. The sample was therefore retrieved by mechanically breaking the assembly using pliers and a chisel. Fig. 7 shows a confocal microscope image of the recovered 10 mm assembly sample, which proves to be an extremely compact solid. During the HPHT treatment, the NaCl pressure-transmitting medium was heated beyond its melting point and mixed with the nanodiamonds, which remained embedded within the volume of the solidified salt upon quenching. The recovered sample also shows that the small platinum spheres remain clearly visible and intact, confirming that the platinum melting temperature was not reached during the experiment, in agreement with the *in situ* calibration.

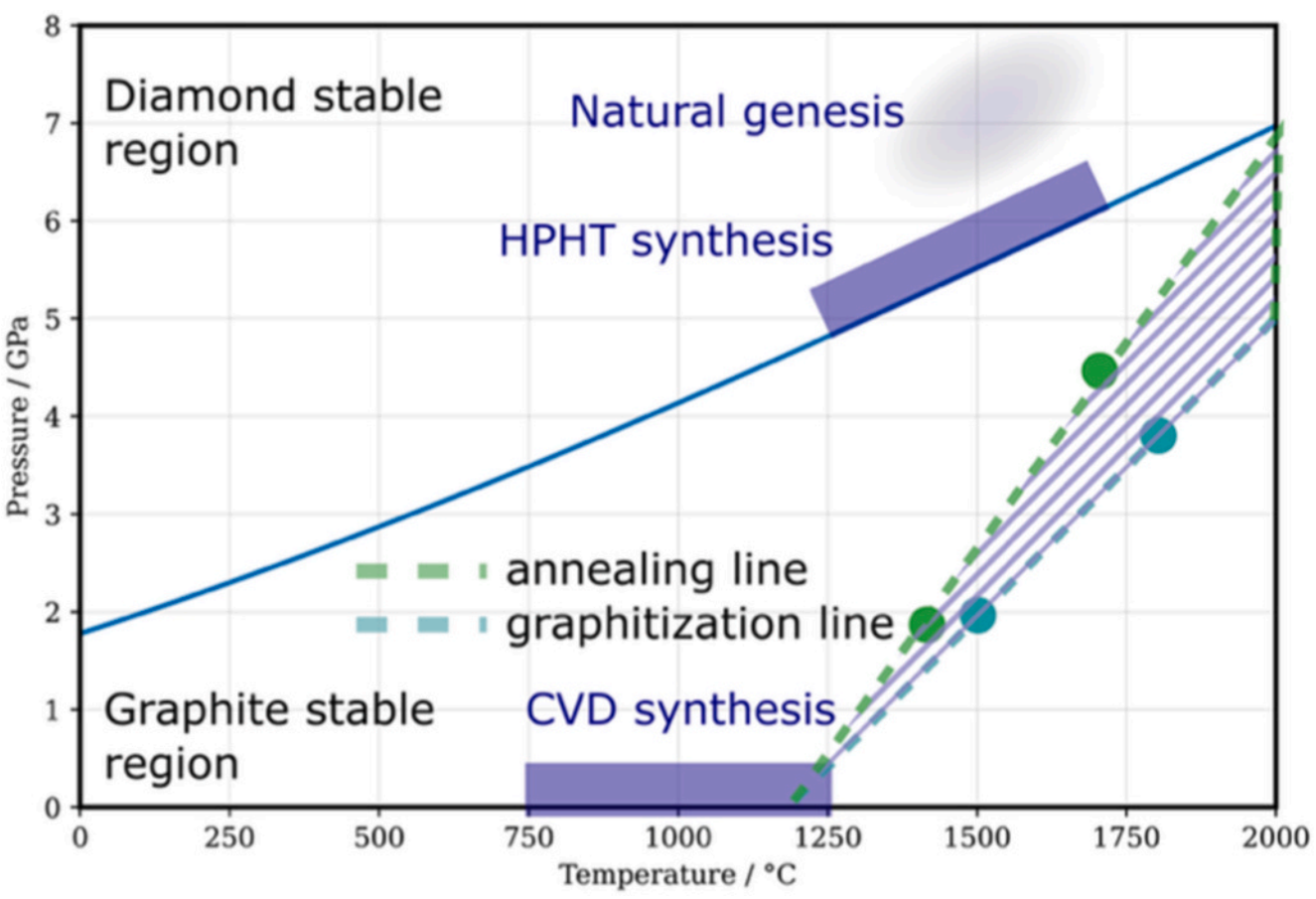

**Fig. 6.** Pressure–temperature diagram of the diamond–graphite phase transition reproduced from Balmer et al. [37], onto which we superimpose our experimental data on the experimentally observed onset of nanodiamond graphitization. Green symbols indicate the experimental annealing conditions (light green) and the onset of graphitization (dark green) determined in this work at 2 and 4 GPa. The shaded region highlights the explored P–T range registered by using NDs powders with characteristic particle sizes of approximately 100–200 nm. (For interpretation of the references to color in this figure legend, the reader is referred to the Web version of this article.)

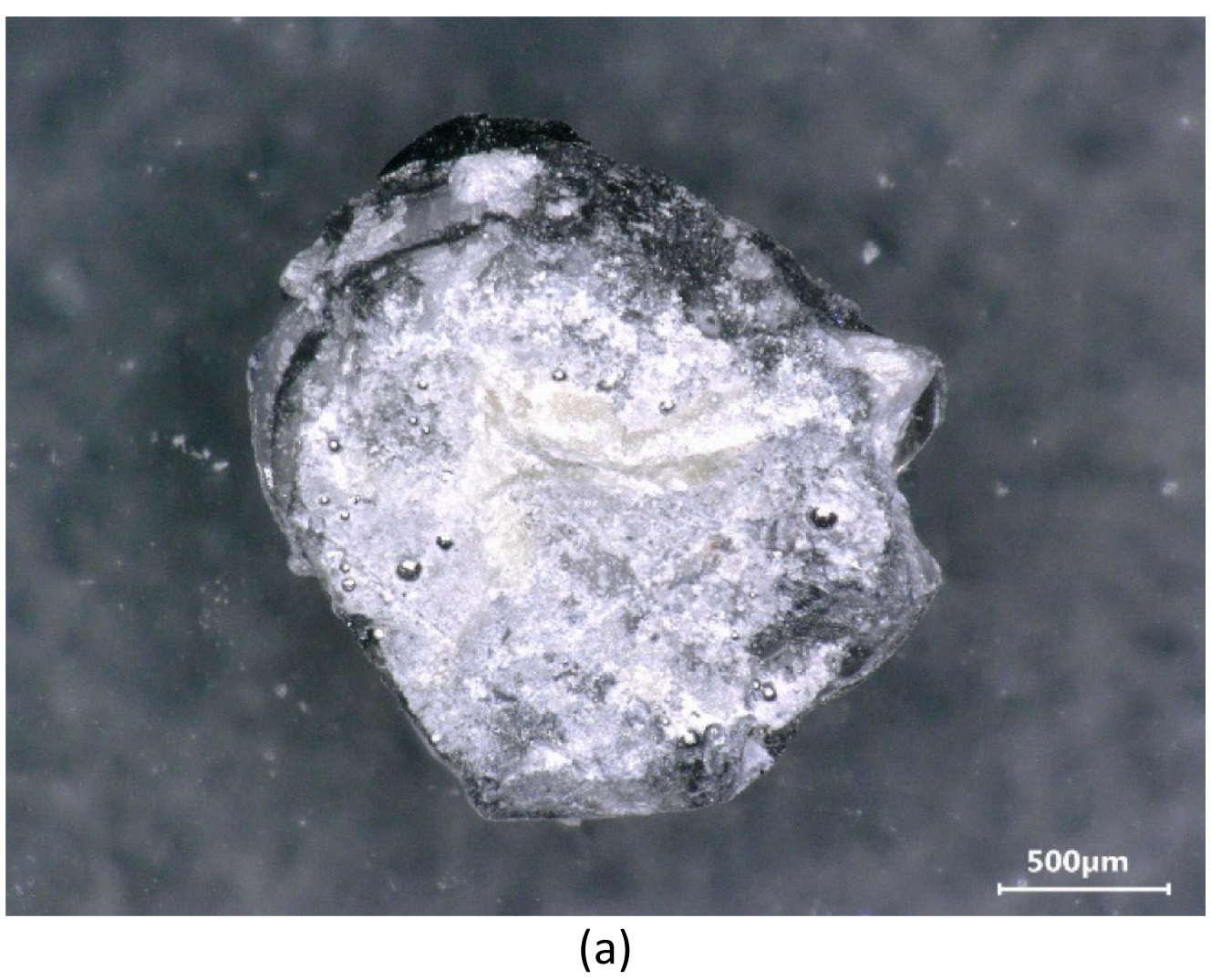

**Fig. 7.** Confocal microscope image of the calibration sample (assembly 10 mm) after annealing treatment in the PE press, once extracted from the assembly. The recovered sample consists of solidified NaCl containing dispersed nanodiamonds together with small silver-coloured platinum spheres.

Subsequently, a dedicated protocol was developed to recover the nanodiamonds from these small salt pieces, which is detailed in the Supplementary Material S4, including SEM and EDS characterizations tracking the process until the salt-free nanodiamonds were obtained. Representative SEM observations of the recovered HPHT-derived nanodiamonds are reported in the Supplementary Material (Fig. S11), showing that the recovered material remains predominantly composed of isolated nanoparticles and small aggregates, with no evidence of extensive particle coalescence following the recovery procedure.

### 3.5. Ex situ structural and optical characterization

Room-temperature and low-temperature photoluminescence measurements were carried out before and after HPHT treatment to evaluate changes in the SiV optical response. Particular attention was paid to linewidth narrowing and to the emergence of partially resolved fine-structure features at cryogenic temperature. Complementary Raman spectroscopy was performed to assess the structural quality of the diamond phase and detect possible graphitic contributions, while SEM was used to examine particle morphology, agglomeration state, and any treatment-induced surface modifications.

A representative SEM image of a recovered CVD-grown SiV-containing ND is shown in Fig. 8a. Fig. 8b illustrates representative Raman spectra collected from three different samples: as-grown nanodiamonds (without any post-treatment), the overheated sample recovered from the 10 mm assembly after exceeding the graphitization threshold, and a new sample of nanodiamonds annealed at 1700 K under 2 GPa for 30 min (selected operating condition, without synchrotron X-ray beam). The as-grown nanodiamonds exhibit the characteristic diamond Raman line at 1332.5 $cm^{-1}$, associated with the first-order Raman scattering of $sp^3$-bonded carbon (FWHM = 6 $cm^{-1}$). The sample annealed at 1700 K retains a sharp diamond peak (same position and FWHM) with no detectable graphitic contribution, indicating preservation of the diamond structure under these HPHT conditions. In contrast, the overheated sample displays a disappearance of the diamond peak in favour of a broad band centred around 1366 $cm^{-1}$, assigned to the D band of graphitic carbon, indicative of $sp^2$ bonding and structural disorder, with its 2D second order at 2732 $cm^{-1}$, and a defined G graphitic peak at 1580 $cm^{-1}$.

The Raman band observed at ~1366 $cm^{-1}$ is attributed to the disorder-induced D mode associated with finite phonon coherence lengths in incompletely graphitized carbon. As shown by Lespade et al. [45], the relaxation of the wave-vector selection rule in nanometric graphitic domains allows non-Γ phonons with high vibrational density of states to contribute to first-order Raman scattering, leading to a D band centred around ~1360 $cm^{-1}$ rather than the canonical ~1350 $cm^{-1}$ of highly ordered graphite. This feature is characteristic of nanocrystalline or turbostratic graphite, where short-range graphitic order dominates despite the absence of long-range three-dimensional stacking.

These spectroscopic observations confirm the *in situ* results, demonstrating that no detectable diamond-to-graphite conversion occurs under the selected pressure–temperature conditions.

Afterwards, in order to optically characterize the SiV color centers, photoluminescence analyses are presented here at room temperature (see Fig. S12 shown in Supplementary Material, S5.) and at 12 K on the following samples (see Fig. 9): as-grown nanodiamonds (without post-synthesis HPHT annealing), nanodiamonds heated at moderate temperatures of 1050 K under 2 GPa pressure named HP1, and nanodiamonds heated at selected temperatures of 1700 K under 2 GPa named HP2.

The as-grown particle spectrum shows the $SiV^-$ ZPL at 739.3 nm with a FWHM of approximately 8.4 nm. The $SiV^-$ ZPL spectra of the HPHT-annealed samples appear very similar to that of the as-grown particles, both in peak position and inhomogeneous broadening, as recorded at room temperature. A small phonon sideband is visible on the right-hand side of the central peak, as expected for measurements at room temperature. These moderate modifications following HPHT annealing highlight the limited sensitivity of room-temperature measurements to strain-related spectral improvements. Low-temperature photoluminescence measurements were therefore performed to more accurately assess the impact of HPHT annealing on the optical properties of the SiV centers.

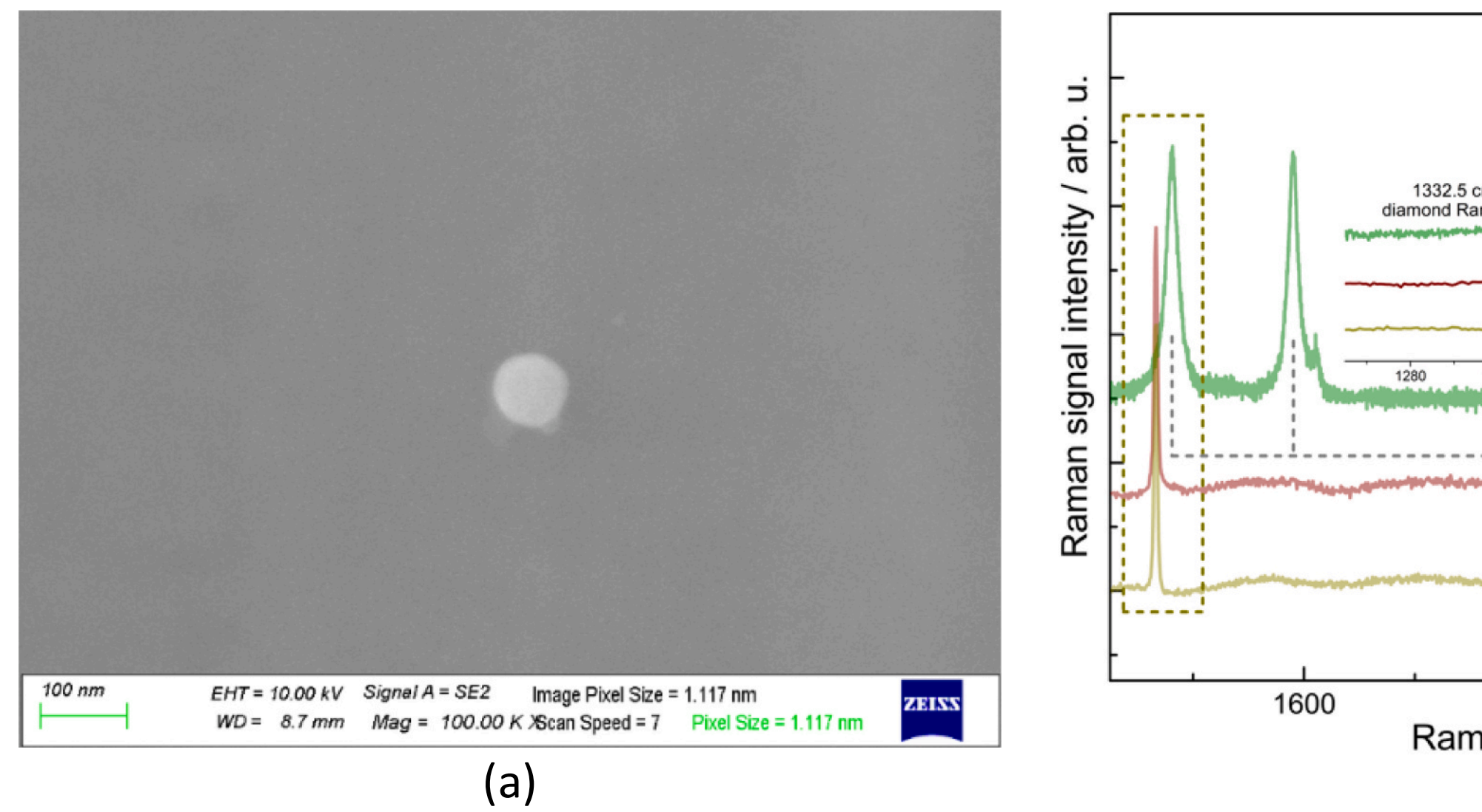

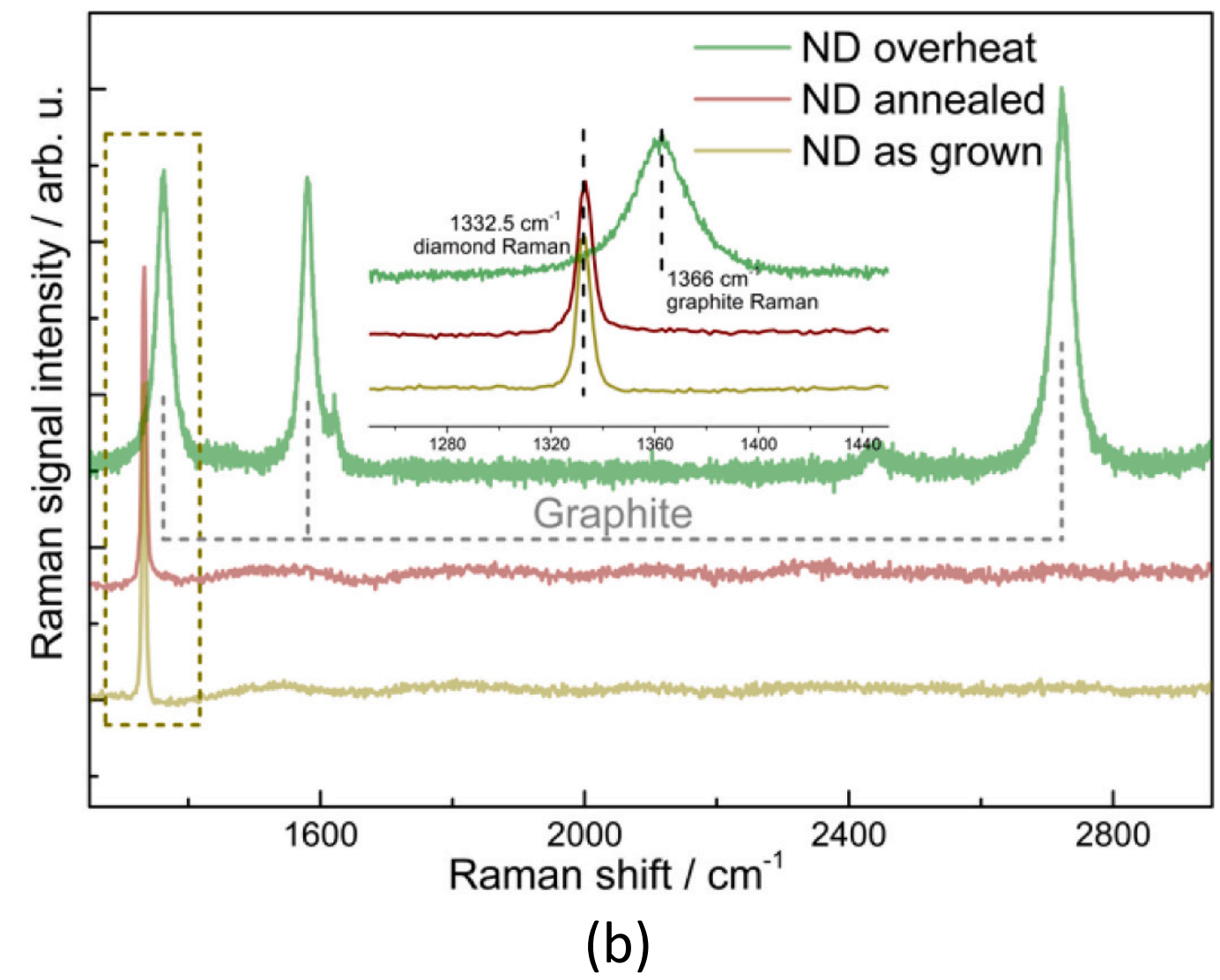

**Fig. 8.** (a) SEM image registered on a CVD-grown SiV-containing ND recovered from the PE assembly after the HPHT annealing treatment (1700 K, under 2 GPa, 30 min). (b) Raman spectra registered on (i) an as-grown ND (yellow line) showing a sharp diamond peak; (ii) an annealed ND (red line) corresponding to the SEM picture (a) and still showing a sharp diamond peak without any graphite traces; (iii) an overheat ND (green line) showing a significant presence of graphite phases at the expense of the diamond one. (For interpretation of the references to color in this figure legend, the reader is referred to the Web version of this article.)

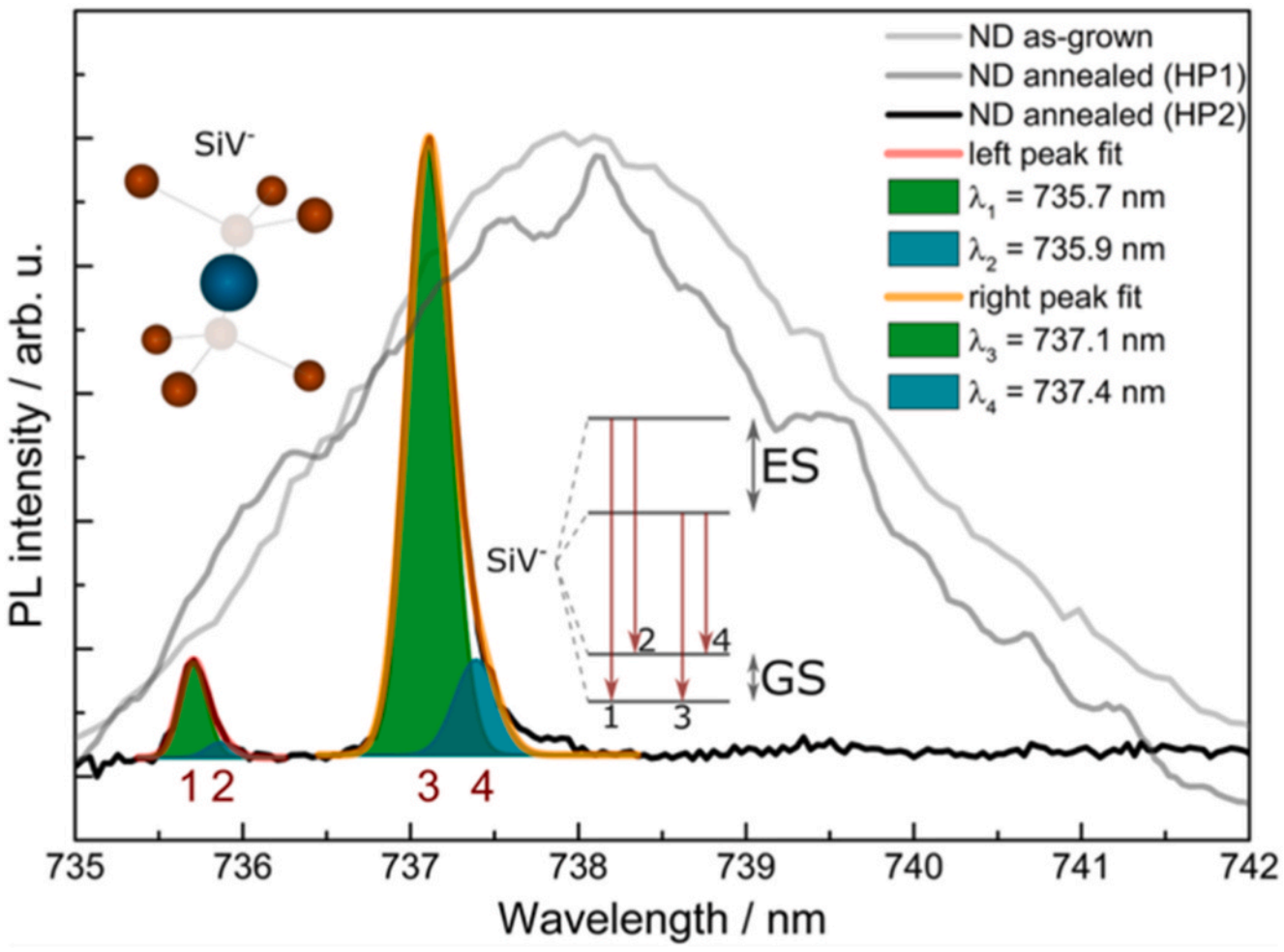

**Fig. 9.** Photoluminescence spectra registered at 12 K of as-grown NDs (light grey line), and HPHT annealed samples (dark grey and black for HP1 and HP2 samples, respectively). Atomic structure of the SiV center and its energy diagram are presented by insets.

In Fig. 9, the SiV PL spectra registered at 12K are presented. The ZPL of the as-grown nanodiamond cluster exhibits a peak at 738.2 nm with a pronounced broadening even at 12 K due to lattice strain (FWHM ≈ 4.3 nm). After annealing under HP1 conditions, the ZPL remains centred at 738.0 nm and still broad preventing access to the fine-structure transitions; nevertheless, a slight narrowing is already observed (FWHM ≈ 4.0 nm), indicating the onset of strain relaxation due to annealing.

In contrast, the spectrum of the HP2-annealed sample shows significant strain relaxation, evidenced by the appearance of two optical transitions characteristic of the $SiV^-$ center fine structure. These two peaks are located at 735.8 nm and 737.2 nm, with FWHM values of 0.1 nm and 0.3 nm, respectively, which are markedly narrower than the 4.3 nm and 4.0 nm values observed for the as-grown and HP1-annealed particles. The corresponding energy separation between the two transitions is approximately 3.2 meV (≈774 GHz). Although the microscopic mechanism responsible for the abrupt improvement observed between the HP1 and HP2 samples cannot be unambiguously identified from the present results, several thermally activated processes may contribute to this behaviour. In particular, the higher annealing temperature reached during the HP2 treatment may promote further relaxation of residual lattice strain, reorganization of crystallographic defects, and reduction of non-radiative recombination centers surrounding the SiV defects. These processes would provide a more homogeneous local crystal environment, thereby reducing inhomogeneous broadening and enabling partial resolution of the SiV fine structure at low temperature. Additional contributions from thermally activated vacancy migration cannot be excluded. The coincidence between the HPHT processing window established by the *in situ* synchrotron X-ray diffraction experiments and the improved low-temperature optical response further suggests that the structural evolution occurring at the highest annealing temperature contributes to the observed optical enhancement, although further investigations will be required to identify the dominant microscopic mechanism.

With regard to the observed spectral splitting, one possible interpretation is that each peak represents the unresolved average of two of the four fine-structure transitions of the $SiV^-$ center, namely 1 + 2 for the 735.8 nm peak and 3 + 4 for the 737.2 nm peak (see the energy level scheme in the figure). Under this assumption, the ground-state splitting would be on the order of ~55 GHz which would be consistent with values reported for low-strain bulk diamond (~50 GHz reported by Müller et al. [2]) although alternative explanations cannot be excluded. However, the inferred excited-state splitting would be ~780 GHz, approximately three times larger than the ~250 GHz typically reported for low-strain bulk diamond ensembles. Such an enhanced splitting indicates a significant perturbation of the excited-state orbital structure, most plausibly induced by local strain. Müller et al. have shown that single SiV centers in nanodiamonds can exhibit substantial spectral shifts (735.5–>737,1 nm at 4 K), attributed to strain-induced modifications of the electronic levels.

In nanodiamonds, optical transitions are more strongly affected than in bulk diamond by surface effects, structural defects, and residual strain. These perturbations can induce both line broadening and shifts in transition energies, leading to a wider spectral distribution of the emitted signals.

Recent first-principles investigations of SiV centers under hydrostatic deformation further demonstrate the pronounced sensitivity of their optical properties to lattice strain [46]. Isotropic compression or tension induces systematic variations of the zero-phonon line and modifies the electronic structure, while sufficiently large deformation may even lead to symmetry lowering from the native $D_3d$ configuration. These results support the interpretation that residual strain fields introduced during high-pressure high-temperature annealing can significantly perturb the orbital structure of the SiV center, potentially accounting for the large excited-state splitting observed here. It should be noted that the spectra shown in Fig. 9 were collected from individual localized nanodiamond particles rather than from ensemble measurements. As each particle may contain multiple SiV centers, the presented spectra are intended to illustrate the improvement in spectral resolution that can be achieved after HPHT annealing, rather than as a statistical description of the entire recovered particle population. Each nanodiamond can contain multiple SiV centers because the synthesis route was specifically designed to produce highly luminescent particles with a high density of color centers [17], suitable for signal recovery under extreme conditions [9]. Consequently, the recorded spectrum represents the collective emission of multiple SiV centers, each possibly experiencing a distinct local strain environment. In this scenario, the two resolved peaks may arise either from strain-enhanced fine-structure splitting or from distinct sub-populations of SiV centers with slightly different zero-phonon line energies. Further lower-temperature measurements with higher spectral and spatial resolution would be required to unambiguously discriminate between these scenarios.

## 4. Conclusions

This work provides a direct experimental determination of the onset of graphitization in nanodiamonds under high-pressure high-temperature conditions. *In situ* synchrotron experiments show that the graphitization onset occurs at 1800 K at 2 GPa and 2120 K at 4 GPa for the applied HPHT heating protocol. These results define a practical HPHT processing window in which nanodiamonds remain structurally stable while undergoing efficient strain relaxation. The established pressure–temperature calibration further enables reliable off-beam HPHT treatments using the same press conditions, providing a practical route for processing nanodiamonds while avoiding graphitization. Within this processing window, individual SiV-containing nanodiamonds exhibit improved optical properties after HPHT annealing, including partial resolution of the $SiV^-$ fine structure at 12 K. This behaviour indicates a reduction in strain-induced broadening in CVD-grown nanodiamonds, consistent with defect recovery and lattice reorganization in diamond materials. More broadly, this work highlights the importance of understanding graphitization during HPHT processing for optimizing the functional properties of color center-containing nanodiamonds. Finally, while the present work demonstrates the feasibility of improving the optical response of SiV-containing nanodiamonds through controlled HPHT annealing, further investigations will be required to evaluate the scalability of this approach across larger nanodiamond populations. The pressure–temperature calibration established here provides a practical basis for these future developments.

## CRediT authorship contribution statement

**M. De Feudis:** Conceptualization, Formal analysis, Funding acquisition, Investigation, Methodology, Project administration, Supervision, Validation, Visualization, Writing – original draft, Writing – review & editing. **B. Yavkin:** Formal analysis, Investigation, Software, Visualization. **L. Henry:** Formal analysis, Investigation, Resources, Writing – review & editing. **K.O. Ho:** Investigation, Software, Visualization. **M.-P. Adam:** Investigation, Methodology, Visualization. **P. Goldner:** Resources, Writing – review & editing. **F. Bénédic:** Investigation, Resources, Writing – review & editing. **S. Desgreniers:** Investigation, Resources, Writing – review & editing. **J.-F. Roch:** Conceptualization, Funding acquisition, Resources, Supervision, Writing – review & editing.

## Declaration of competing interest

The authors declare that they have no known competing financial interests or personal relationships that could have appeared to influence the work reported in this paper.

## Acknowledgements

The authors gratefully acknowledge Carla Bittencourt for performing the XPS analyses and for her invaluable scientific discussions and interpretation of the surface chemistry results; Gregory Lefèvre for his collaboration on the DLS measurements, his valuable advice on particle size analysis, and many insightful scientific discussions; and Maria Konstantakopoulou for her expertise in scanning electron microscopy, X-ray diffraction analyses, and valuable scientific discussions. The authors also thank Jean-Sébastien Lauret for his assistance with low-temperature spectroscopy measurements; Nicolas Guignot for his guidance during the experiments conducted at Synchrotron SOLEIL; Savino Longo for valuable discussions on the thermodynamic and kinetic aspects of graphitization under HPHT conditions; Jocelyn Achard and Huan-Cheng Chang for fruitful scientific discussions on CVD-grown diamond materials and HPHT-synthetised nanodiamond properties, respectively. Finally, the authors acknowledge Vanna Pugliese and Noémie Pham for their assistance with experimental spectrometry and physicochemical analyses, respectively, as well as Lingraj Kumar and Dillan Belhadji for their contributions to the graphical abstract, DLS analyses, and XRD figure preparation, respectively.

This work was supported by the French Agence Nationale de la Recherche (ANR) through the *NanoG4V* project under Grant Agreement No. ANR-24-CE51-7558 (PI: Mary De Feudis). Mary De Feudis and Serge Desgreniers acknowledge the support of the Mourou–Strickland Mobility Grant between France and Canada (2023), funded by the Embassy of France in Canada through the project "*Nanodiamonds: CVD synthesis and color centers incorporation for quantum technologies*", which supported the scientific collaboration between their laboratories and contributed to the present work. The authors also acknowledge support from the ANR through the ESR/EquipEx + program (Grant No. ANR-21-ESRE-0031), and the European Research Council (ERC) under the Advanced Grant *QPRESSE* (Grant No. 101142682), PI: Jean-François Roch. The QuanTip regional network on quantum technologies of Île-de-France region is also gratefully acknowledged for its financial support. Jean- François Roch acknowledges support from Institut Universitaire de France.

## Appendix A. Supplementary data

Supplementary data to this article can be found online at https://doi.org/10.1016/j.carbon.2026.121960.